# WILI – WEB INTERFACE FOR PEOPLE WITH LOW-VISION ISSUES

K.S.Kuppusamy[1], Leena Mary Francis[2] and G.Aghila[3]

[1]Department of Computer Science, School of Engineering and Technology, Pondicherry University, Pondicherry, India
`kskuppu@gmail.com`

[2]Department of Computer Science, SS College, Pondicherry, India
`leena.francis.pdy@gmail.com`

[3]Department of Computer Science, School of Engineering and Technology, Pondicherry University, Pondicherry, India
`aghilaa@yahoo.com`

## ABSTRACT

*Though World Wide Web is the single largest source of information, it is ill-equipped to serve the people with vision related problems. With the prolific increase in the interest to make the web accessible to all sections of the society, solving this accessibility problem becomes mandatory. This paper presents a technique for making web pages accessible for people with low vision issues. A model for making web pages accessible, WILI (Web Interface for people with Low-vision Issues) has been proposed. The approach followed in this work is to automatically replace the existing display style of a web page with a new skin following the guidelines given by Clear Print Booklet provided by Royal National Institute of Blind. "Single Click Solution" is one of the primary advantages provided by WILI. A prototype using the WILI model is implemented and various experiments are conducted. The results of experiments conducted on WILI indicate 82% effective conversion rate.*

## KEYWORDS

*Web accessibility, WILI, low vision issues*

## 1. INTRODUCTION

Today's World Wide Web is highly interactive and designed with advanced technologies to increase communication and collaboration [1]. At the same time user interface is complex, lacks intuitiveness and it is not easy to be used by people with visual challenges [2]. Overlaid with a Graphical user interface (GUI), visually impaired users find the web increasingly difficult to access, navigate and interpret [3]. A model is proposed for making the web pages accessible, named WILI (Web Interface for people with Low-vision Issues; pronounced as "Vizhi" meaning eye in Tamizh language).

This paper concentrates on users having low vision. When a person cannot be corrected to normal 20/20 vision with ordinary glasses, he or she is said to have "low vision"[4]. This condition can result from inheritance, trauma, disease, or aging. People with low vision still have useful vision that can often be improved with visual devices [5].

The main features of this model are as given below:

- The approach is to automatically replace the existing display style of a web page with a new skin by following the guidelines given by Clear Print Booklet[2].

- Providing a "Single Click Solution" feature which is one of the primary advantages of WILI model.

The remainder of this paper is organized as follows. In Section 2, motivations for this research are provided. Section 3 deals with the architecture of the proposed system and its structured representation. Section 4 explains the experimental setup. Results of experiments are provided in Section 5. Conclusions and future enhancements are given in Section 5.

## 2. MOTIVATIONS

There are different ways for personalizing a web page's presentation such as accessibility features of standard browsers, transcoding proxy servers, application adaptors/mediators, zooming features and special purpose browsers [6].

Standard browsers allow users to select color schemes or font sizes and types that suit the user, and override the design dictated by the HTML codes. This does require some configuration of the browser, which may not be obvious to less experienced users, especially the more complex changes (e.g., customizing a Cascaded Style Sheet file to match one's preference) [7]. Few browsers do support screen readers but rich and multimedia content is increasing rapidly on the Web [8]. It is very attractive for sighted people, but it brings severe problems to screen reader users. Once the audio starts playing, it becomes hard for the visually impaired users to listen to the screen reader [9].

If the proxy server changes the HTML codes provided by the web server en route back to the user, it is referred to as transcoding proxy server [10]. This appears to offer a reasonable way to present information according to a user's needs, but they do have some significant drawbacks. The most significant drawback is that it may not support page features, such as client redirects, and many websites assume the direct use of a client and provide functionality based on this assumption (e.g., the use of cookies to track users and provide password-authenticated services to them) [11].

An application adaptor or mediator is positioned in front of the browser between the browser and the user [12]. It performs the transformation of the HTML document after parsing by the browser. The drawback is that this requires installation of the mediator on the local machine, which may not be possible for the user, as not all users have administrator privilege or feel confident installing the software [13].

Assistive technologies can be simply installed according to their requirements e.g., a screen magnifier or reader to enlarge and read out screen information, text summarization software to simplify the text [14]. The problem with traditional accessibility techniques is that they were designed to handle a single disability. Secondly, some accessibility techniques products are difficult to learn and use [15].

All the above factors led to an improvised model named as WILI (Web Interface for people with Low-vision Issues) in which the user directly types in the URL and by single click the page is retrieved, modified and given back to them as accessible web page. And also the user will be able to navigate to other pages through links of the accessible web page.

## 3. WILI ARCHITECTURE

The idea of WILI is to provide accessible web interface to low vision users. The overall design of WILI is as shown in Fig. 1. The architecture of WILI includes two major components namely, Style parser and Link parser.

- Style Parser – this parses the style element, identifies the equivalent style (ES) and replace the source style with ES.

- Link Parser – this parses the link element, checks whether the nature of URL is relative or absolute and if absolute it removes the base portion.

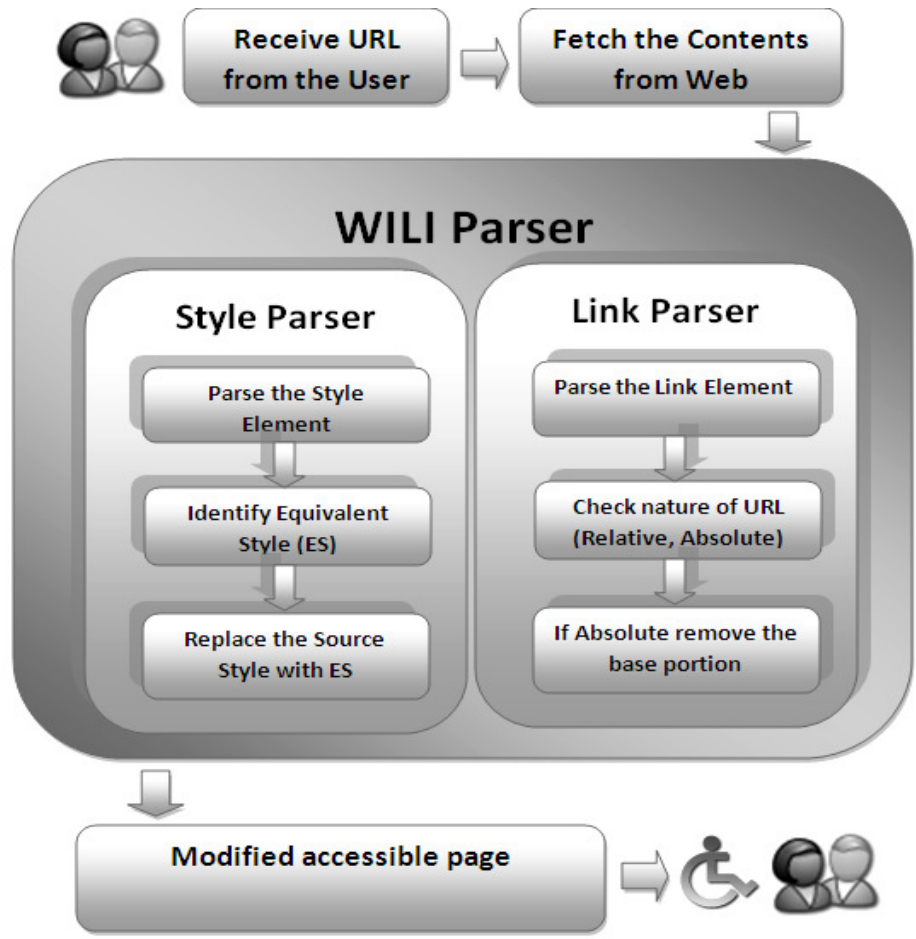

Figure 1. WILI Architecture

### 3.1. WILI Model

The proposed system receives the page URL from the user and then fetches the required contents from the Web.

$$\Omega = \omega[\lambda] \tag{1}$$

International Journal on Computational Sciences & Applications (IJCSA) Vo2, No.2, April 2012

In (1), $\lambda$ represent the URL and $\omega$ represent the fetching function from the World Wide Web. The fetched content is parsed by both of the parsers as shown in (2).

$$\Omega = \{\Omega^S, \Omega^L\} \qquad (2)$$

The Style parser is responsible for the change of style which makes the page accessible to the users. It parses the style element, identifies the equivalent style (ES) and replaces the source style with ES. The new style followed by accessible web pages are the rules specified in the Clear Print Booklet.

$$\Omega^S = \{\forall_{i=1..n} \Omega^S(i) = \Gamma(i)\} \qquad (3)$$

The Link parser is responsible for the navigation of the user to other pages through links, wherein the navigated pages also becomes accessible web pages to the user.

$$\Omega^L = \{\forall_{i=1..n} \Omega^L(i) = \Psi[\Omega^L(i)]\} \qquad (4)$$

### 3.2. WILI Algorithm

The algorithmic representation of the above said procedure is as shown below:

>    Step 1: Receive the URL of the source page, SURL, as input.
>    Step 2: Fetch the contents of the page pointed by SURL as C [ SURL ].
>    Step 3: Pre-process the contents of C [ SURL ].
>        Step 3.1: Extract the display style components in S [ SURL ] as ST.
>            Step 3.1.1: For each element ti in ST :
>                Step 3.1.1.1: Identify the equivalent accessible style EST[ti] for ti.
>                    Step 3.1.1.1: Delete ti from C [ SURL ].
>                    Step 3.1.1.2 : Insert EST[ti]into C [ SURL ].
>        Step 3.2: Extract the link components in C [ SURL ] as L.
>            Step 3.2.1: For each li in L:
>                Step 3.2.1.1 : Parse the link element li.
>                Step 3.2.1.2 : If RELATIVE[li]  GOTO Step 3.2.1.4
>                Step 3.2.1.3 : Remove the base component b[li] from li.
>                Step 3.2.1.4 : Make the link element li point to WILI server.
>    Step 4: Send back the contents of C[SURL] to the client.

## 4. EXPERIMENTS

In order to evaluate the performance of the proposed system, a prototype implementation has been developed. The technologies used for this implementation are Hypertext Preprocessor (PHP) on server side, JavaScript and HTML on the client side. Screen shots of WILI implementation are shown below. The user interface which accepts the URL from the user is shown in Fig 2. The contents are retrieved and parsed. The accessible web page is shown in Fig 3. And the original web page is shown in Fig 4.

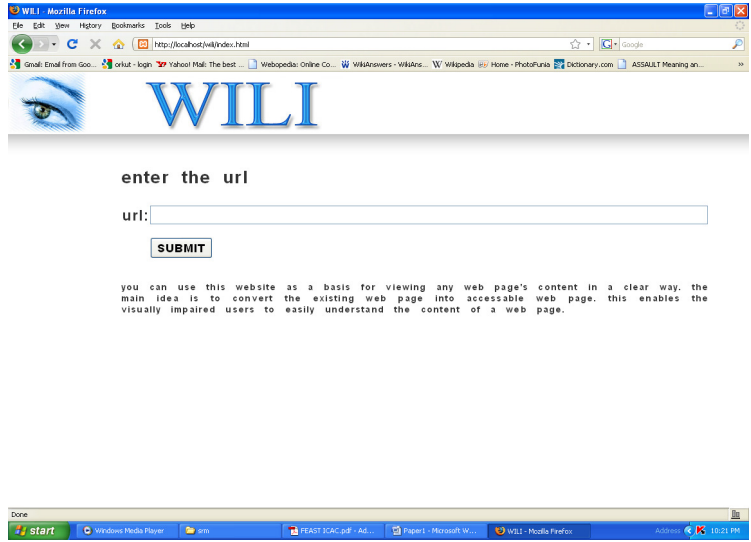

Figure 2. WILI Home Page

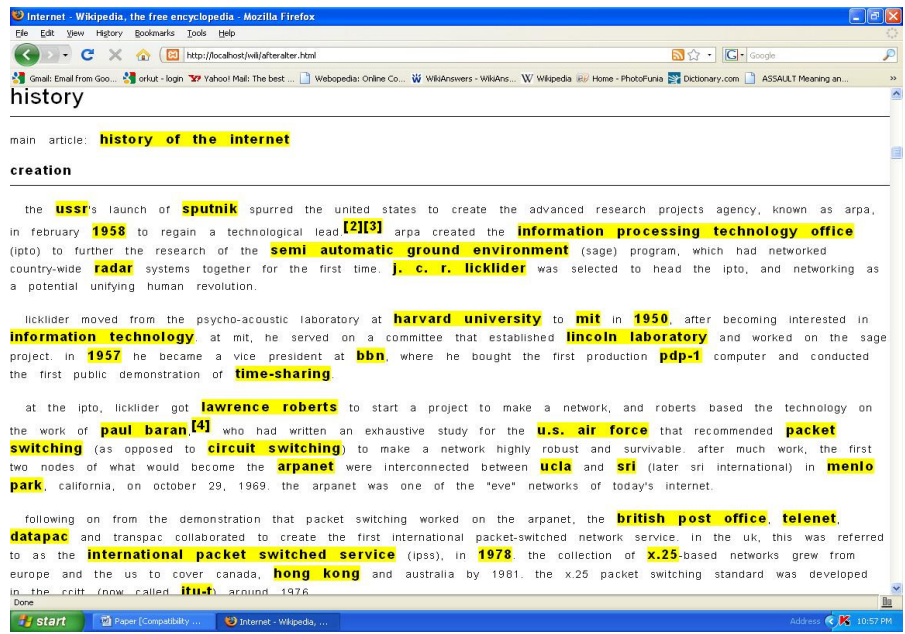

Figure 3. Accessible Page

International Journal on Computational Sciences & Applications (IJCSA) Vo2, No.2, April 2012

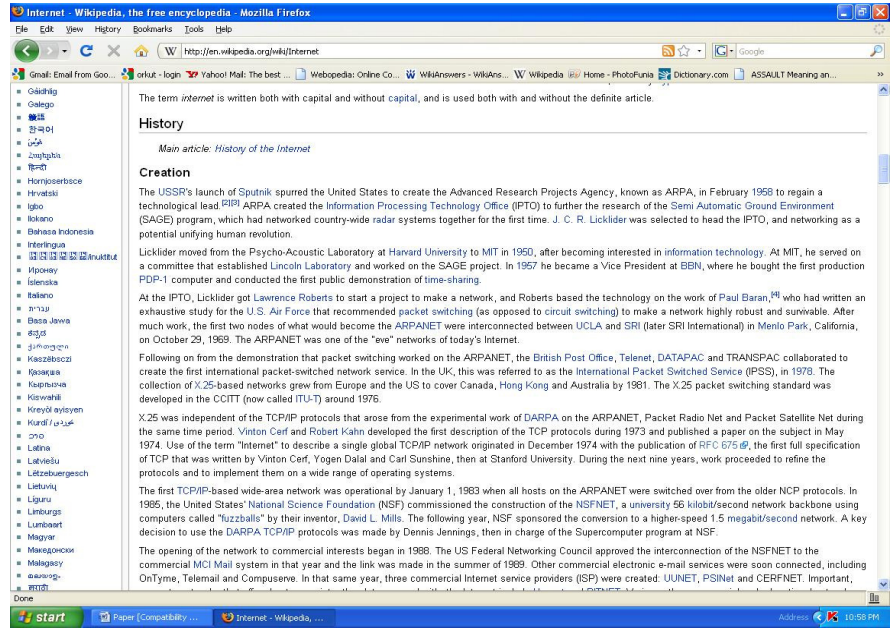

Figure 4. Source Web Page

## 5. RESULT ANALYSIS

Few experiments were conducted on the prototype implementation described in the previous section. For the experiment ten batches [B1 to B10] of URLs were taken into consideration. These batches of URLs were tested for load time and rate of conversion of the original web page into accessible page. Table 1 shows comparison of execution time of web page taken normally versus WILI and the rate of conversion. WILI takes lesser time than the usual time taken by the URL to load, as it overrides the complex style and scripts embedded in the web page which is proved by Fig 5. WILI was tested for similarity of the text content between the original page and accessible page which is measured in percentages. Fig. 6 shows that conversion rate doesn't deviate too much and has produced 82% effective conversion rate.

Table 1. Load time and conversion rate

| Batch | NLT | WLT | Conversion Rate |
|---|---|---|---|
| B1 | 3.85 | 1.71 | 79 |
| B2 | 6.7 | 3.4 | 63 |
| B3 | 7.91 | 4.97 | 86 |
| B4 | 10.81 | 4.68 | 84 |
| B5 | 15.88 | 10.75 | 96 |
| B6 | 7.42 | 3.82 | 85 |
| B7 | 9.48 | 7.91 | 86 |
| B8 | 13.08 | 5.07 | 78 |
| B9 | 17.6 | 6.37 | 81 |
| B10 | 10.16 | 2.78 | 79 |

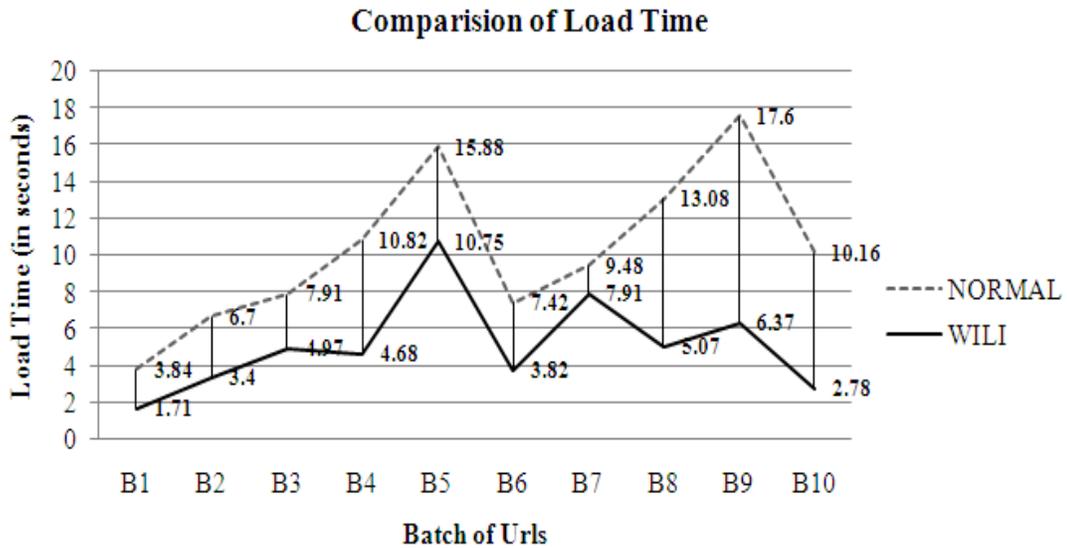

Figure. 5 Comparison of load time

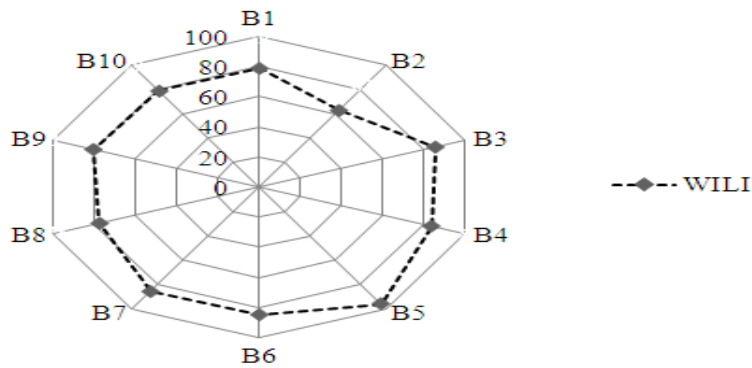

Fig. 6 Conversion rate

## 6. CONCLUSIONS

The experiments conducted on current implementation are encouraging. Web pages from various sources were given as output and WILI produced 82% effective conversion rates. The current implementation of WILI holds only the essential features required for making the pages accessible to users with low vision issues. The WILI model can be enhanced by including various features like user-specific style sheets etc. The pictures embedded in web page can be zoomed into bigger size when the mouse pointer moved over it and the content display can be shown in much more refined format. The regional language support of WILI model can also be enriched.

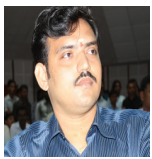
K.S.Kuppusamy is an Assistant Professor at Department of Computer Science, School of Engineering and Technology, Pondicherry University, Pondicherry, India. He has obtained his Masters degree in Computer Science and Information Technology from Madurai Kamaraj University. He is currently pursuing his Ph.D in the field of Intelligent Information Management. His research interest includes Web Search Engines, Semantic Web. He has made 11 international publications.

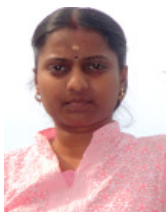
Leena Mary Francis is an Assistant Professor at Department of Computer Science, SS College, Pondicherry, India. She has obtained her Masters degree in Computer Applications from Pondicherry University. Her area of interest includes Web 2.0 and Information retrieval. She has worked at Oracle before joining the current position.

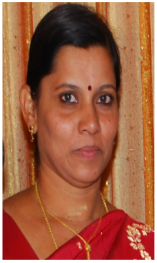
G. Aghila is a Professor at Department of Computer Science, School of Engineering and Technology, Pondicherry University, Pondicherry, India. She has got a total of 22 years of teaching experience. She has received her M.E (Computer Science and Engineering) and Ph.D. from Anna University, Chennai, India. She has published more than 55 research papers in web crawlers, ontology based information retrieval. She is currently a supervisor guiding 8 Ph.D. scholars. She was in receipt of Schrneiger award. She is an expert in ontology development. Her area of interest includes Intelligent Information Management, artificial intelligence, text mining and semantic web technologies.